\documentclass[preprint,aps,tightenlines,floatfix,showpacs]{revtex4}

\usepackage{graphics}
\usepackage{bm}
\usepackage{amsmath}
\usepackage{amssymb}
\begin{document}

\title{Predicting proton-nucleus total reaction cross sections up to 300 MeV 
using  a simple functional form}
\author{P. K. Deb}
\email{pkd@physics.unimelb.edu.au}
\affiliation{School of Physics, The University of Melbourne, Victoria
3010, Australia}
\author{K. Amos}
\email{amos@physics.unimelb.edu.au}
\affiliation{School of Physics, The University of Melbourne, Victoria
3010, Australia}

\date{\today}
\begin{abstract}
Total reaction cross sections are predicted for proton scattering from
various nuclei.  A simple functional form has been used  that reproduces 
the total reaction cross sections for the scattering of protons from (15)
nuclei spanning the mass range from ${}^9$Be to ${}^{238}$U and for proton 
energies 10 to 300 MeV. 
\end{abstract}
\maketitle

\section{Introduction}

Reaction cross sections from the scattering of nucleons by nuclei
(stable and radioactive) are required in a number of fields of study;
some being of quite current interest~\cite{De01}. Those fields range  
from problems in basic science through many of applied nature.
An example of the latter is the
transmutation of long lived radioactive waste into shorter lived
products using accelerator driven systems (ADS). 
As well, proton-nucleus
($p$-$A$) cross section values at energies to 300 MeV or more are needed
not only to specify important quantities of relevance to proton 
radiation therapy~\cite{Jo01}, but also as they are key
information in assessing radiation protection for patients.
As an example in basic science, total reaction cross sections are important
ingredients to a number of problems in astrophysics.
Also the integral observable of 
proton scattering from a given nucleus gives
direct information on the neutron root mean square radius in 
nuclei~\cite{Bro2000}; a
property sought in parity-violating electron scattering
experiments~\cite{Je00}.

It would be very utilitarian if aspects of such scattering
were well approximated by a simple convenient
functional form.  Recently it has been shown~\cite{Amos02a}
that, for neutron total reaction cross sections, such a form may exist.
Herein we consider that concept further to reproduce the 
measured total reaction cross sections from 
proton scattering for  energies  ranging from
10 MeV to 300 MeV, and  from  15 nuclei
ranging in mass from 9 to 238.

\section{ Formalism}

The total reaction cross sections for proton scattering 
from nuclei can be expressed in terms of partial wave
scattering ($S$) matrices specified 
at energies $E\propto k^2$, by
\begin{equation}
S^{\pm}_l \equiv S^{\pm}_l(k) = e^{2i\delta^{\pm}_l(k)} =
\eta^{\pm}_l(k)e^{2i\Re\left[ \delta^{\pm}_l(k) \right] }\ ,
\end{equation}
where $\delta^\pm_l(k)$ are the (complex) scattering
phase shifts and $\eta^{\pm}_l(k)$ are the moduli
of the $S$ matrices. The superscript designates
$j = l\pm 1/2$.
Total reaction cross sections then follow from 
\begin{eqnarray}
\sigma_R(E)  &=&  \frac{\pi}{k^2} \sum^{\infty}_{l=0} \left\{ (l+1)
\left[ 1 - \left( \eta^+_l \right)^2 \right] + l \left[ 1 - \left(
\eta^-_l \right)^2 \right] \right\}\nonumber\\
&=& \frac{\pi}{k^2}\sum^{\infty}_{l=0} \sigma_l^{(R)}(E)\ .
\label{xxxx}
\end{eqnarray}
Despite the successes with use of the  $g$-folding approach
to define $NA$ optical potentials~\cite{Review}, we
note that there are discrepancies between the predictions 
of total reaction cross sections found thereby and 
actual data. Those usually are due to limitations 
with the structure
model used to describe the ground state densities of some nuclei. However,
there have been so many successes with the approach 
~\cite{Review,De01,Kara02,Amos02} that we believe the functional form developed
here on the basis of matching the values obtained with g-folding potentials
are  pertinent initial guesses to start a
refinement (of the parameter values) to reproduce actual measured
total reaction cross sections.

As evident from the figures presented in a recent paper~\cite{Amos02a},
the partial total reaction
cross sections, $\sigma_l^{(R)}(E)$,
can be described by the simple functional form,
\begin{equation}
\sigma_l^{(R)}(E) = (2l+1) \left[1 + e^{\frac{(l-l_0)}{a}}\right]^{-1}
 + \epsilon (2l_0 + 1)
 e^{\frac{(l-l_0)}{a}}
\left[ 1 + e^{\frac{(l-l_0)}{a}} \right]^{-2}
\label{Fnform}
\end{equation}
with $l_0(E,A)$, $a(E,A)$, and $\epsilon(E,A)$ varying smoothly with
energy and mass.

The summation giving the total reaction cross section
can be limited to a value $l_{max}$ and the associated
form tends appropriately to the
known high energy limit.  With increasing energy,
$l_{max}$ becomes so large that  the
exponential fall  of the functional form, Eq.~\ref{Fnform},
can be approximated as a straight vertical line
($l_0 = l_{max}$).
In that case, the total reaction cross section
equates to the area of a triangle, and
\begin{equation}
\sigma_R \Rightarrow \frac{\pi}{2 k^2} l_{max}(2l_{max} + 1)
 \approx \frac{\pi}{k^2} l_{max}^2\ .
\end{equation}
Then with $l_{max} \sim kR$, at high energies
\begin{equation}
\sigma_R \Rightarrow \pi R^2\ ;
\end{equation}
the geometric cross section as required.


\section{ Results and discussions}

Although on using Eqs.~\ref{xxxx} and \ref{Fnform} to match values of
(theoretically) calculated total reaction cross sections led~\cite{Amos02a} 
to the three parameters 
$l_0(E,A)$, $a(E,A)$, and $\epsilon(E,A)$  having smooth variations 
with energy, there are discrepancies between those predictions 
and the actual measured data.
Herein we modify the method of selection of those parameter values 
to produce more accurate reaction cross section values, while 
keeping as smooth a variation with energy of those parameters as
possible. 
Specifically, in Eq.~\ref{Fnform}, we
have set the $\epsilon$ as a constant ($-1.5$) and so independent of energy 
and of mass. Further we assume that $a(E,A)$ varies linearly
with the wave vector, 
\begin{equation}
k = \frac{1}{\hbar c} \sqrt{E^2 - m^2 c^4}\ ,
\end{equation}
and with the form
\begin{equation}
a(E,A)
\sim 1.02 k - 0.25\ . 
\label{fn-values}
\end{equation}
Then $l_0(E,A)$ were adjusted to ensure that all measured total reaction
cross section values are matched by using the function form, Eq.~\ref{Fnform}.
\begin{figure}
\scalebox{0.9}{\includegraphics{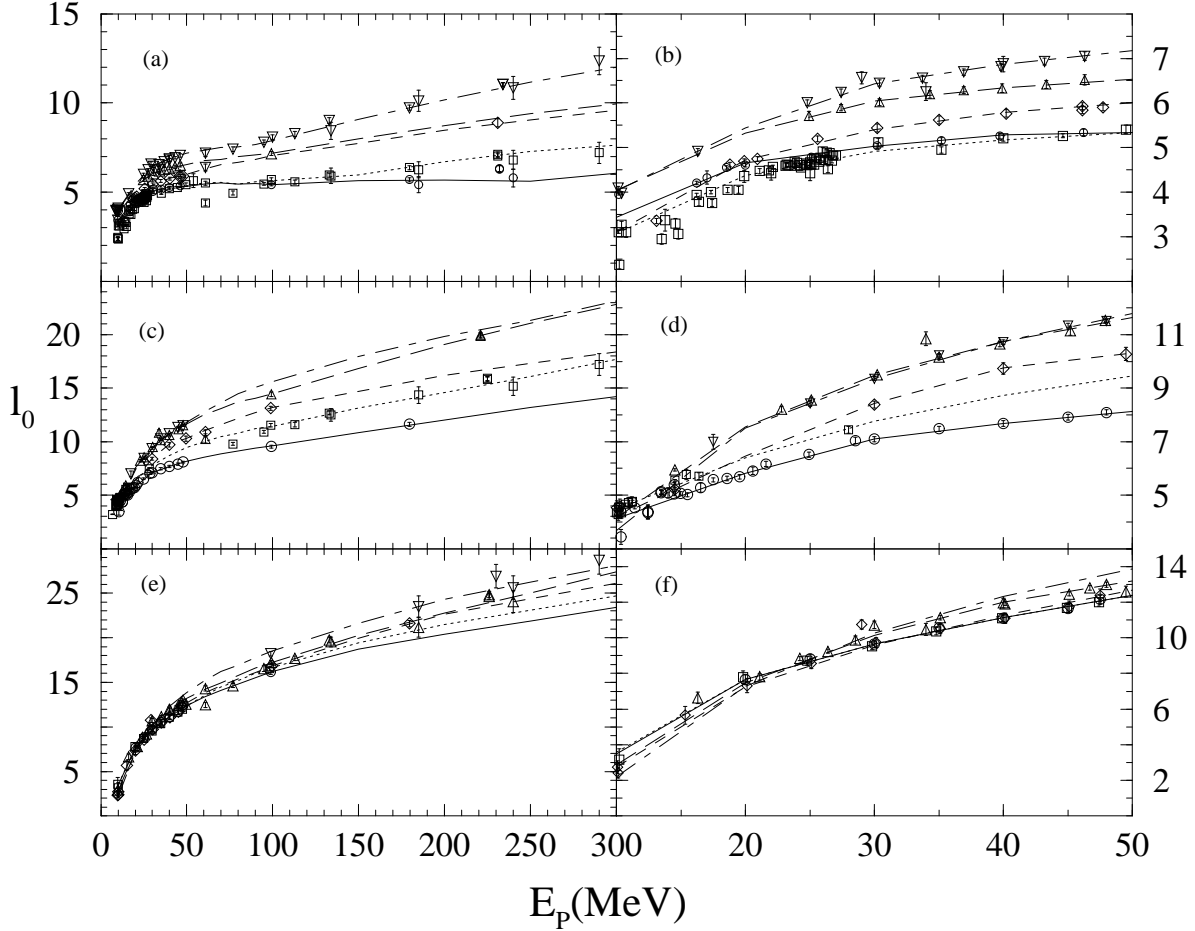}}
\caption{\label{l0-energy}
The parameter values for $l_0$ in the simple functional form for the 
$p$-$A$ reaction cross sections. Points shown with error bars are the 
values required to match actual measured data values.}
\end{figure}

The resultant optimized values for the 
parameter $l_0$ are presented in Fig.~\ref{l0-energy}.
In this figure, the points represent the calculated values of $l_0$ 
at particular energies where the experimental values of total reaction 
cross sections are available for the considered nuclei. 
We have used the lower and upper uncertainty 
limits to get the error bars for $l_0$ values. We call these $l_0$ values 
with uncertainties as "data-$l_0$" hereafter. The curves result on
using a spline interpolation on those data points constrained
to give an optimal smoothness variation of $l_0(E,A)$ with E.
The figure displays the results for the entire energy range (to 300 MeV)
in the left hand panels, while the right hand panels emphasize the 
low energy variations to 50 MeV since most experimental values lie
in that range of energy.

In panel (a) of Fig.~\ref{l0-energy},  
the solid, dotted, dashed, long-dashed, and dot-dashed lines depict the 
values for $^9$Be, $^{12}$C,  $^{16}$O, $^{19}$F, and $^{27}$Al respectively, 
as do circles, squares, diamonds, up triangles and down triangles for 
the data-$l_0$ results. The same legends apply to the 10 to 50 MeV
plots shown in panel (b).
In panels (c) and (d),  the energy variations of $l_0$ values for $^{40}$Ca,
$^{63}$Cu, $^{90}$Zr, $^{118}$Sn, and $^{140}$Ce are presented by solid, 
dotted, dashed, long-dashed and dot-dashed lines respectively. The 
corresponding data-$l_0$ points are also presented by circles, squares,
 diamonds, up triangles and down triangles. 
Finally , in panels (e) and (f), we display
the variations of $l_0$ with energy for $^{159}$Tb,
$^{181}$Ta, $^{197}$Au, $^{208}$Pb and $^{238}$U by solid, dotted, dashed, 
long-dashed and dot-dashed lines respectively with the
corresponding data-$l_0$ points presented by circles, squares,
diamonds, up triangles and down triangles.

Numerical values for $l_0$ that result from the simple functional form 
calculations and as displayed by the curves in Fig.~\ref{l0-energy}, are listed 
at various energies in Table.~\ref{lvalues}. For energies 50 MeV and higher,
these $l_0$ values monotonically increase with mass and energy as might
be anticipated. At lower energies however, and while the $l_0$ values
do still vary monotonically with energy, there is structure  in their mass
variation. In particular,
the values of $l_0$ for $^{197}$Au, $^{208}$Pb and $^{238}$U in the 
energy range from 10 MeV to 20 MeV decrease with mass. 
But note that there is little data available for these nuclei at this
energy range. 
\begin{table}
\begin{ruledtabular}
\caption{\label{lvalues}
$l_0$ values for different nuclei at different energies.} 
\begin{tabular}{cccccccccccccccc}
  \multicolumn{1}{c}{Energy} & \multicolumn{15}{c}{Nucleus} \\
(MeV) & ${}^9$Be & $^{12}$C &  $^{16}$O & $^{19}$F & $^{27}$Al & $^{40}$Ca &
$^{63}$Cu & $^{90}$Zr & $^{118}$Sn & $^{140}$Ce & $^{159}$Tb & $^{181}$Ta &
$^{197}$Au & $^{208}$Pb & $^{238}$U \\
\hline
 10 & 3.43 & 3.08 & 3.11 & 4.04 & 4.02 & 4.12 & 4.47 & 4.18 & 4.15 & 
3.67 & 3.47 & 3.57 & 2.75 & 2.87 & 2.19 \\
20 & 4.65 & 4.37 & 4.72 & 5.32 & 5.44 & 5.82 & 6.40 & 6.46 & 7.56 & 7.51 & 7.65 & 7.66 & 7.24 & 7.45 & 7.28 \\
30 & 5.03 & 4.91 & 5.40 & 6.04 & 6.43 & 7.09 & 7.77 & 8.42 & 9.47 & 9.36 & 9.67 & 9.65 & 9.60 & 10.14 & 10.26 \\
40 & 5.29 & 5.17 & 5.78 & 6.34 & 6.86 & 7.69 & 8.73 & 9.76 & 10.74 & 10.75 & 
11.13 & 11.13 & 11.25 & 11.99 & 12.31 \\
50 & 5.34 & 5.34 & 6.00 & 6.53 & 7.17 & 8.14 & 9.46 & 10.31 & 11.64 & 11.79 & 
12.37 & 12.38 & 12.68 & 13.19 & 13.87 \\
60 & 5.44 & 5.50 & 6.28 & 6.75 & 7.35 & 8.56 & 10.01 & 10.98 & 12.50 & 12.76 &
13.34 & 13.48 & 13.75 & 14.18 & 15.15 \\
70 & 5.51 & 5.54 & 6.53 & 6.83 & 7.49 & 8.88 & 10.46 & 11.60 & 13.10 & 13.65 & 
14.12 & 14.47 & 14.61 & 14.94 & 16.22 \\
80 & 5.46 & 5.45 & 6.66 & 7.00 & 7.52 & 9.13 & 10.80 & 12.16 & 13.76 & 14.46 & 
14.83 & 15.22 & 15.41 & 15.76 & 16.90 \\
90 & 5.48 & 5.57 & 6.85 & 7.08 & 7.66 & 9.38 & 11.19 & 12.71 & 14.04 & 15.06 &
15.53 & 16.02 & 16.09 & 16.59 & 17.79 \\
100 & 5.42 & 5.64 & 7.05 & 7.19 & 7.88 & 9.63 & 11.44 & 13.20 & 14.48 & 15.61 & 16.21 & 16.64 & 16.88 & 17.25 & 18.51 \\
150 & 5.63 & 5.97 & 7.72 & 7.98 & 9.09 & 10.87 & 13.14 & 14.70 & 16.82 & 17.95 & 18.71 & 19.41 & 19.92 & 20.25 & 21.65 \\
200 & 5.65 & 6.70 & 8.47 & 8.73 & 10.17 & 12.04 & 14.58 & 16.18 & 19.08 & 19.86
& 20.68 & 21.50 & 22.23 & 22.75 & 24.34 \\
250 & 5.61 & 7.29 & 9.06 & 9.39 & 11.15 & 13.17 & 16.12 & 17.90 & 20.19 & 21.33
 &  22.38 & 23.65 & 24.40 & 25.01 & 26.24 \\
300 & 6.07 & 7.64 & 9.58 & 9.96 & 12.01 & 14.23 & 17.22 & 20.05 & 22.37 & 23.79
 & 24.70 & 25.91 & 26.90 & 27.41 & 28.11 \\
\end{tabular}
\end{ruledtabular}
\end{table}

\begin{figure}
\scalebox{1.0}{\includegraphics{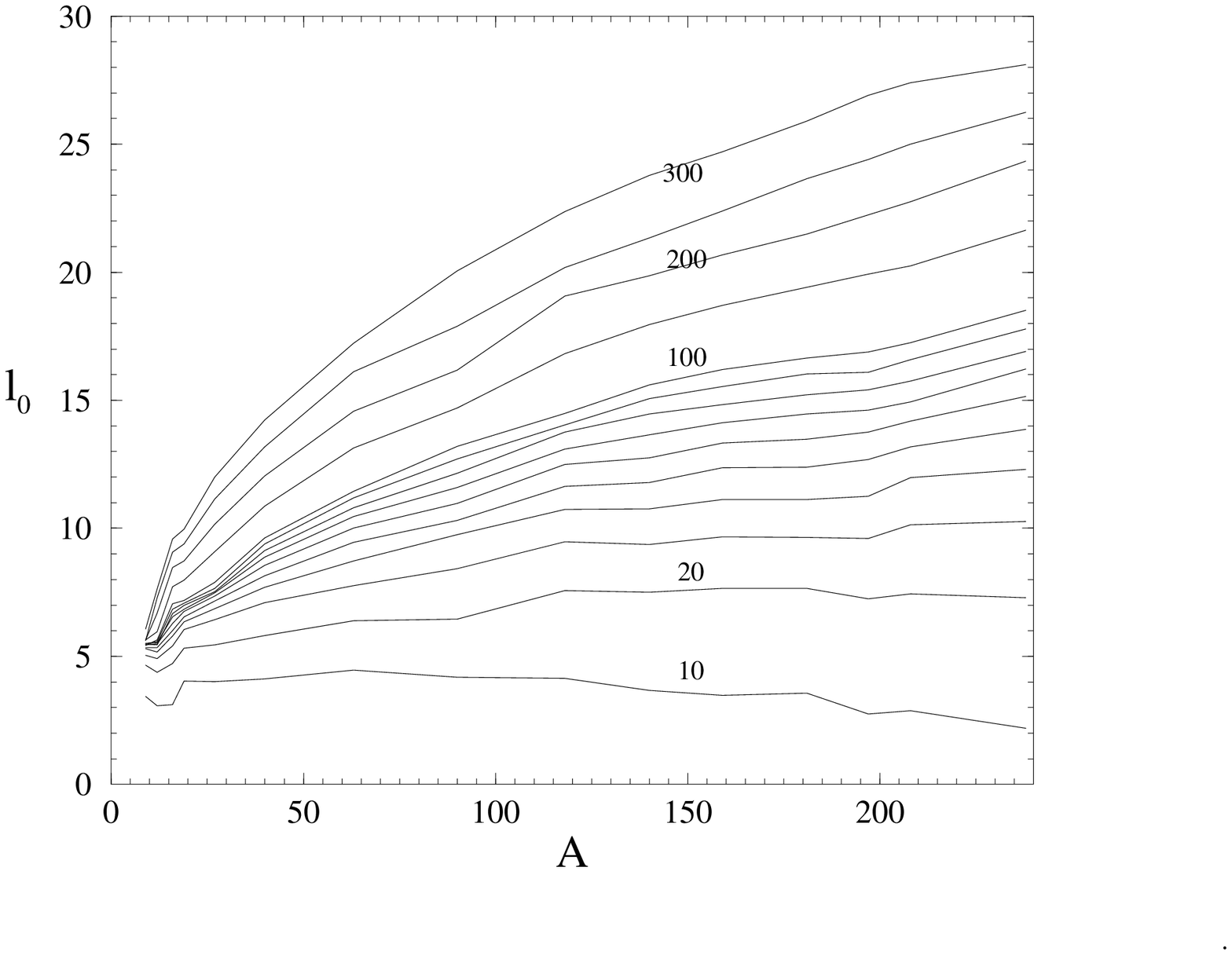}}
\caption{\label{mass-plot}
The parameter values for $l_0$ in the simple functional form for the 
$p$-$A$ reaction cross sections as a function of target mass for each
energy given in the table.}
\end{figure}

The mass variations given by these tabled values have extra structure
as is evident from the plots given in Fig.~\ref{mass-plot}.
The results are not as smooth as the counterpart variation with energy
particularly at the lowest of the energy set.
That may reflect the scattered (in energy) nature of the actual data 
from which starting values of $l_0$ have been taken to generate the functional forms.

\begin{figure}
\scalebox{0.9}{\includegraphics{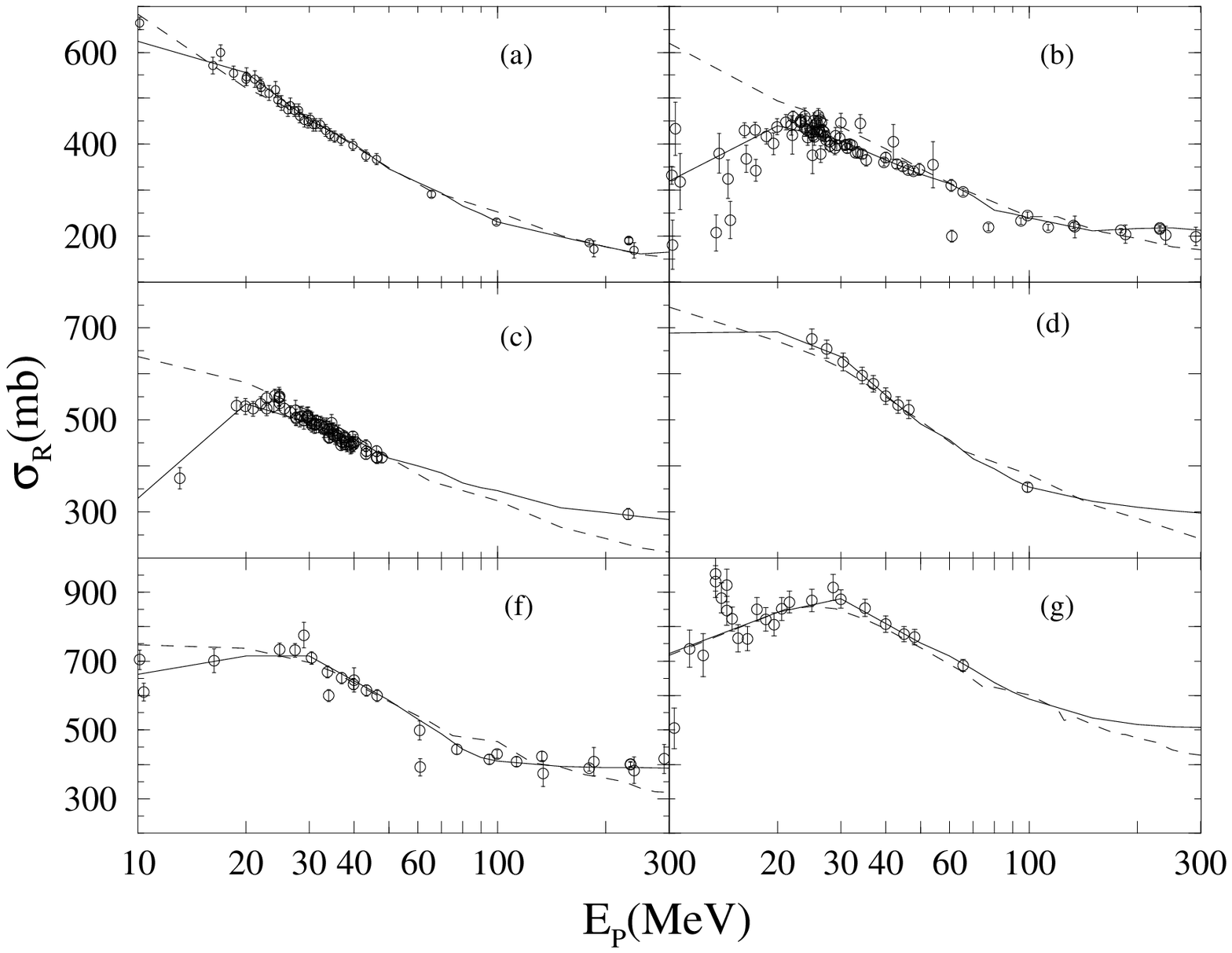}}
\caption{\label{9-40-sigma}
Energy dependence of $\sigma _R$ for proton scattering from 
(a) $^9$Be, (b) $^{12}$C, (c) $^{16}$O, (d) $^{19}$F, (e) $^{27}$Al and 
(f) $^{40}$Ca. Dashed lines represent the results obtained from $g$-folding
optical potential calculations while the solid curves portray
the values obtained by using simple functional form.}
\end{figure}

\begin{figure}
\scalebox{0.9}{\includegraphics{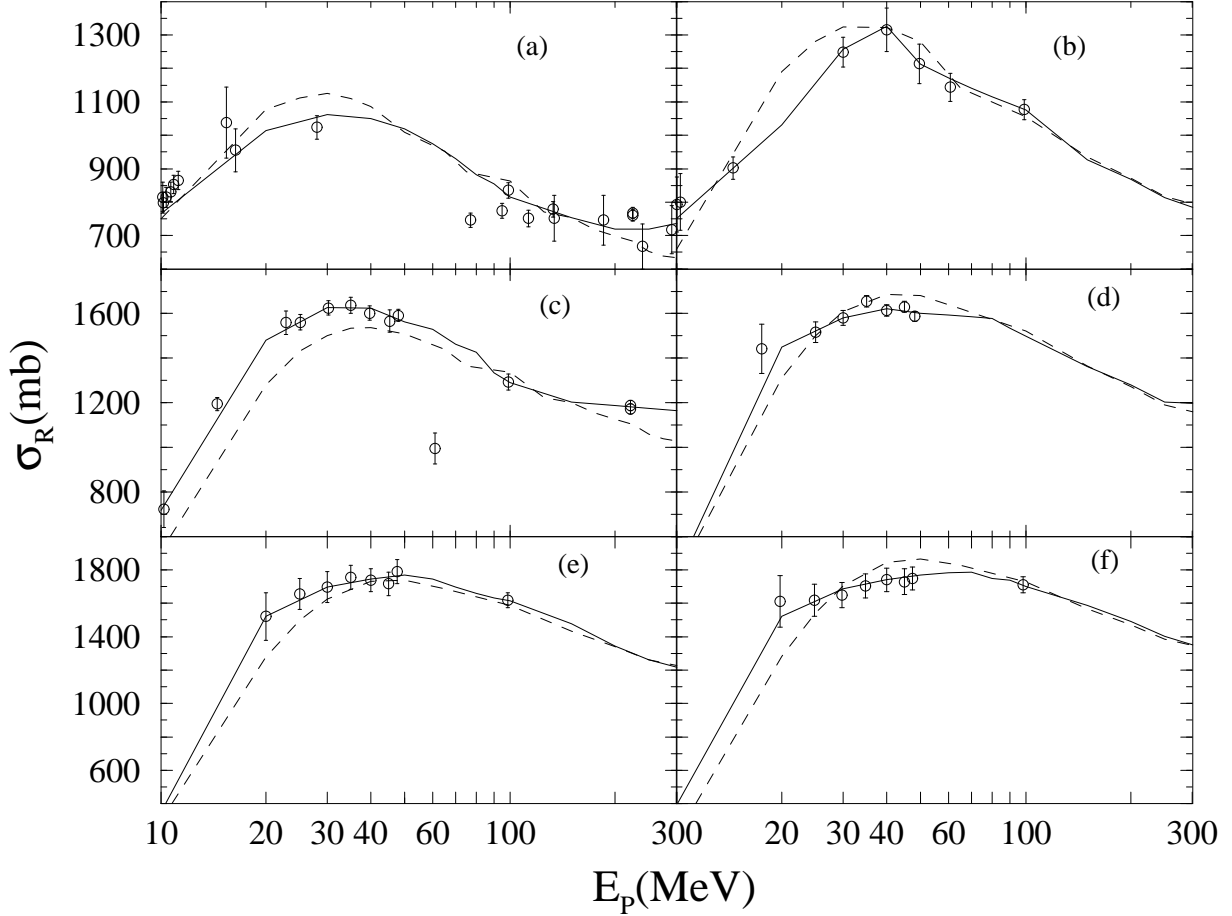}}
\caption{\label{63-181-sigma}
Same as Fig.~\ref{9-40-sigma} but from 
(a) $^{63}$Cu, (b) $^{90}$Zr, (c) $^{118}$Sn, (d) $^{140}$Ce, 
(e) $^{159}$Tb and 
(f) $^{181}$Ta.}
\end{figure}

\begin{figure}
\scalebox{0.9}{\includegraphics{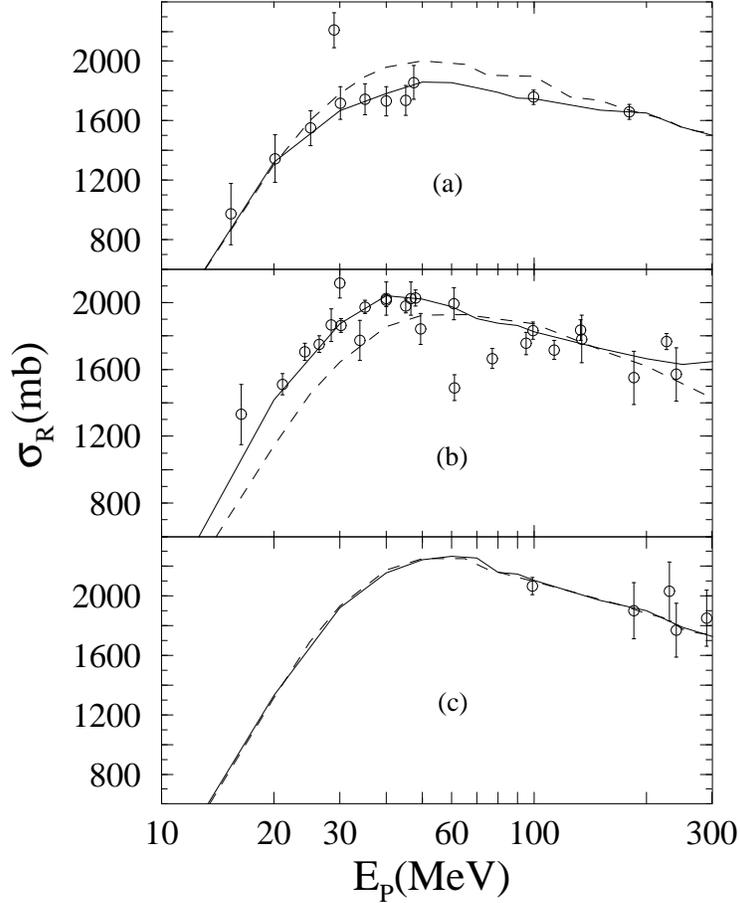}}
\caption{\label{197-238-sigma}
Same as Fig.~\ref{9-40-sigma} but from
(a) $^{197}$Au, (b) $^{208}$Pb, and (c) $^{238}$U.}
\end{figure}

In Figs.~\ref{9-40-sigma},~\ref{63-181-sigma}, and ~\ref{197-238-sigma},
we show the total reaction cross sections generated using the simple 
functional form and tabled values of $l_0$, and displayed by the solid 
curves, with those obtained from calculations made using $g$-folding optical 
potentials~\cite{Amos02}.  Dashed lines represent the predictions obtained from
those microscopic optical model calculations. The experimental 
data~\cite{Car96} are depicted by circles.   

The results for scattering from $^9$Be, $^{12}$C, $^{16}$O, $^{19}$F, $^{27}$Al,
and $^{40}$Ca are displayed in segments (a) through (f) 
of Fig.~\ref{9-40-sigma} respectively.
In segment (a), the data are well reproduced by the $g$-folding 
predictions resulting from the folding with the $^9$Be ground state OBDME 
found with (0+2)$\hbar\omega$ spectroscopy as well as
 by those obtained with the simple functional form method. 

Calculated p-$^{12}$C reaction cross sections are compared with the 
experimental data in segment (b) of Fig.~\ref{9-40-sigma}. The reaction cross
sections obtained from $g$-folding calculations are in good 
agreement with the experimental data but only in the energy range above 20 MeV.
On the other hand, the results obtained from the simple functional 
method are excellent for all energies, replicating the average trend of data
well even in the lower energy range from 10 MeV. There are
two data points, at 61 MeV and at 77 MeV, in disagreement with the 
calculated results however. But, as noted previously~\cite{De01},
these data points should be discounted. 

Predictions for p-$^{16}$O and for p-$^{19}$F scattering are compared 
with data in segments (c) and (d) of Fig.~\ref{9-40-sigma}.
For p-$^{16}$O case, there are many
data points at the energies between 20 to 40 MeV. Predictions from $g$-folding
calculations while replicating the data well at and above 25 MeV,
overestimate at lower energies. That $g$-folding result also underestimates
the datum at 250 MeV; the sole datum above 50 MeV. In contrast, and by design,
the results obtained from 
the simple functional method  are in excellent agreement with the experimental
data at all energies. For p-$^{19}$F, although $g$-folding calculations
reproduce the data  well, the simple functional form method gives slightly 
more accurate predictions.

Total reaction cross section
predictions for p-$^{27}$Al and for p-$^{40}$Ca  
are compared with the experimental data in segments (f) and (g) of 
Fig.~\ref{9-40-sigma}. Again while  $g$-folding calculations reproduce the data 
quite well to 200 MeV, three data points between 180 and 300 MeV are not 
matched.
The $g$-folding results underestimate them noticeably. But predictions from 
a simple functional form approach replicate the data well at all 
energies. One data point at 61 MeV is exceptional in the set.
With $^{40}$Ca, the folding model approach is not expected to be reliable 
at the energies in the range 10 to 20 MeV, as is the case with $^{12}$C,
since for excitation energies of that range, both nuclei have 
clearly discrete spectra.  That is true for most light mass nuclei but little 
or no total reaction cross sections have been reported for them.  Indeed the 
reaction data from both $^{12}$C and from $^{40}$Ca show rather sharp 
resonance features below 20 MeV. Both the $g$-folding calculations and 
functional form calculations reproduce the rest of the $^{40}$Ca data well. 

The results for scattering from $^{63}$Cu, $^{90}$Zr, $^{118}$Sn, $^{140}$Ce,
$^{159}$Tb, and $^{181}$Ta are displayed in Fig.~\ref{63-181-sigma} in segments 
(a) through (f) respectively.  
For  $^{63}$Cu, predictions at low energies may be slightly 
too small and the parameter sets driven too severely by the sole datum
at 30 MeV in the range 20 to 70 MeV.  Also the data in the range 100
to 300 MeV are quite scattered but the simple functional form gives
a good average result.

In segment (b) of Fig.~\ref{63-181-sigma}, the predicted total reaction cross
sections from p-$^{90}$Zr scattering are compared with the experimental data.
Results from $g$-folding calculations are in good agreement with  the 
data although the data value at 30 MeV is overestimated. 
The results obtained from 
the simple functional form, and by dint of construction,
 are in excellent agreement with the experimental 
data at the few energies measured, but the shape is not optimally smooth.
This we believe is the prime cause for the kink shown at $A = 90$ in the 
mass variation plots in Fig.~\ref{mass-plot} and most noticeably at 200 MeV.
 Lack of data meant that we had to use the $g$-folding 
values to specify the functional form.  That is also the case with masses 
140, 159, and 181.

The p-$^{118}$Sn total reaction cross section results are given in segment (c)
of Fig.~\ref{63-181-sigma} where the two predictions again
are compared with the data.
Although not as good as the results found for scattering from light mass 
nuclei, the $g$-folding potential still gives reasonable shape prediction. 
However, the model underestimates the data by 5 to 10\%. But  
the simple functional 
form  model form again gives an excellent reproduction of the data 
at all energies except 
61 MeV. This 61 MeV data point is again exceptional being much smaller
than other data and the predictions as in the cases of $^{12}$C and 
$^{27}$Al.

Predictions for p-$^{140}$Ce, p-$^{159}$Tb and for p-$^{181}$Ta scattering 
are compared with the (limited amount of) data in segments (d), (e) and (f) 
of 
Fig.~\ref{63-181-sigma} respectively. The $g$-folding calculations give 
good agreement with that data for p-$^{140}$Ce, slightly underestimate
the data for p-$^{159}$Tb, and for p-$^{181}$Ta,
underestimate data at the energies to 20 MeV and overestimate data in the 
energy range 40 to 60 MeV. In all cases results predicted by the simple
functional form are  excellent reproductions of the experimental data.

In segments (a), (b) and (c) of Fig.~\ref{197-238-sigma} we compare the 
calculated total reaction cross sections for p-$^{197}$Au, p-$^{208}$Pb and
p-$^{238}$U scattering with the available data. 
For $^{197}$Au, the $g$-folding optical potential calculations
are in good agreement with most data; the 29 MeV datum grossly 
underestimated by the calculations. But that data point is also at odds with 
the energy trend of the other data. Save for that 29 MeV value,
the simple functional form gives even better predictions of the data.  
The energy variation of the p-$^{208}$Pb reaction cross sections is shown 
in segment (b) of Fig.~\ref{197-238-sigma} where the predictions
from $g$-folding optical potential calculations and from the
simple functional form 
calculations are compared with a fairly  extensive set of experimental data. 
In making the $g$-folding potentials, 
we have used Skyrme-Hartree-Fock  wave functions~\cite{Bro2000} which 
have been shown to be more
realistic~\cite{Amos02} than  simple oscillator model ones. 
Still such  $g$-folding calculations 
underestimate the data up to 50 MeV. But simple functional form calculations
are in excellent agreement with the experimental data, save that
data values at 30, 61 and 77 MeV again are exceptional. 
The predictions of total reaction cross sections for p-$^{238}$U scattering 
from the $g$-folding optical potential calculations and from using the simple 
functional form are compared with the few data
in segment (c) of Fig.~\ref{197-238-sigma}.  Given the lack of data
the two results are virtually identical.

As a final note regarding many of the exceptional data values so defined
in the foregoing, Menet {\it et al.}~\cite{Men71} argue  that there may be 
a systematic error in the studies reported in those experiments.

\section{ conclusions}

The measured reaction cross sections for 10--300 MeV proton scattering
from nuclei ranging in mass from $^9$Be to $^{238}$U are well reproduced
by calculations made using a three parameter function. The values 
of the parameters ($l_0$, $a$, $\epsilon$) are set in a simple manner and
when set by enough actual data,
can be used to predict the total reaction cross sections at any 
energy for a given  nucleus. 
The mass variations of those parameters at fixed energies also
are smooth, and very much so when data exists to control the results,
and we believe that the simple functional form also can be used to
find reasonable estimates of the total reaction cross sections
of protons from any stable nucleus in the mass range.


\begin{acknowledgments}
This research was supported by a research grant from the Australian
Research Council.
\end{acknowledgments}

\end{document}